# ArchiveGPT:

## A human-centered evaluation of using a vision language model for image cataloguing

Line Abele[1], Gerrit Anders[2], Tolgahan Aydın[2], Jürgen Buder[2], Helen Fischer[2], Dominik Kimmel[3], Markus Huff[1, 2]

[1] Department of Psychology, University of Tübingen, Germany

[2] Leibniz-Institut für Wissensmedien, Tübingen, Germany

[3] Leibniz-Zentrum für Archäologie, Mainz, Germany

**Abstract**

The accelerating growth of photographic collections has outpaced manual cataloguing, motivating the use of vision language models (VLMs) to automate metadata generation. This study examines whether AI-generated catalogue descriptions can approximate human-written quality and how generative AI might integrate into cataloguing workflows in archival and museum collections. A VLM (*InternVL2*) generated catalogue descriptions for photographic prints on labelled cardboard mounts with archaeological content, evaluated by archive and archaeology experts and non-experts in a human-centered, experimental framework. Participants classified descriptions as AI-generated or expert-written, rated quality, and reported willingness to use and trust in AI tools. Classification performance was above chance level, with both groups underestimating their ability to detect AI-generated descriptions. OCR errors and hallucinations limited perceived quality, yet descriptions rated higher in accuracy and usefulness were harder to classify, suggesting that human review is necessary to ensure the accuracy and quality of catalogue descriptions generated by the out-of-the-box model, particularly in specialized domains like archaeological cataloguing. Experts showed lower willingness to adopt AI tools, emphasizing concerns on preservation responsibility over technical performance. These findings advocate for a collaborative approach where AI supports draft generation but remains subordinate to human verification, ensuring alignment with curatorial values (e.g., provenance, transparency). The successful integration of this approach depends not only on technical advancements, such as domain-specific fine-tuning, but even more on establishing trust among professionals, which could both be fostered through a transparent and explainable AI pipeline.

**Introduction**

Photographic archives and collections preserve vast stores of visual historical, art historical, and archaeological evidence, yet their accelerating growth has outpaced the manual cataloguing on which scholarly access still depends. Vision language models (VLMs) promise to automate description and metadata generation, but whether their output satisfies professional expectations of accuracy, provenance, and contextual richness is unclear. Using the photographic holdings of the Leibniz-Zentrum für Archäologie (LEIZA) as a use case, we explored whether a VLM can draft catalogue entries that experts will trust, use, and perhaps even mistake for human work. This article speaks to both the technical and sociocultural stakes of bringing generative AI into practice in gallery, library, archive, and museum (GLAM) institutions.

A primary challenge lies in the transformation of archives into data-driven infrastructures. Millions of digitized photographs circulate as research assets whose value remains dormant behind incomplete metadata; human curators must now collaborate with computational agents to manage records across the archival continuum (Colavizza et al., 2021). Recent progress in VLMs has produced single architectures that analyze images and generate coherent descriptive prose (Muehlberger et al., 2019). Federated approaches even combine several models to improve consistency and reliability, outperforming single-model solutions in large-scale tests (Groppe et al., 2025). Complementary work in digital libraries shows that computational pipelines can cut error rates in automatic metadata generation to near-negligible levels (Karnani et al., 2022).

Technical promise alone is insufficient. Archivists scrutinize descriptions through the prisms of authenticity, provenance, and ethical stewardship, and recent work warns that AI will not be taken up unless those values remain visible and negotiable (Jaillant & Rees, 2023). Behavioral evidence reinforces the point: across two experiments with representative German samples, the balance of perceived risks and opportunities was the strongest predictor of people's willingness to use an AI system, overruling demographics and even domain expertise (Schwesig et al., 2023). Follow-up analyses show that when the context evokes higher perceived risk (e.g., transport over medicine), adoption drops, even where technical performance is unchanged, highlighting how contextualised trust thresholds govern uptake (Schwesig et al., 2023). Crucially, archival professionals often regard the insertion of inaccurate or misleading metadata as a high-impact risk. Even minor factual errors can propagate through scholarly networks and undermine the evidential value of entire collections (Bunn, 2020; Gilliland, 2012).

Cognitive factors deepen this trust equation. Large-scale studies demonstrate that greater topic knowledge can induce "risk blindness", leading users to underestimate hazards, whereas high metacognitive confidence amplifies perceived benefits; both distortions sway acceptance of AI outputs unless counterbalanced by transparent explanation and critical literacy. These findings echo large-scale reviews showing public trust in AI hinges on clear explanation and professional literacy (Arias Hernández & Rockembach, 2025; Kleizen et al., 2023). Consequently, archival debates have expanded, discussing not only the question "can the system perform?" but also the more pressing "will professionals and publics accept its results, and on what terms?" (Davet et al., 2023).

**Research questions and hypotheses**

Building on the challenges outlined above, this study examines how generative AI, specifically a VLM, can be meaningfully integrated into cataloguing photographic materials in archives and collections. Central to this investigation is the question of whether AI-generated catalogue descriptions can match or approximate the quality, usefulness, and perceived trustworthiness of those written by human experts like professional archivists and collection managers. In contrast to prior studies that focus primarily on technical performance or metadata extraction (e.g., Fischer et al., submitted; Muehlberger et al., 2019), this study adopts a human-centered experimental framework, assessing users' perceptions of AI-generated content and their reactions regarding trust and acceptance. The psychological experiment involved both experts (in archiving and/or archaeology and lay people, as both groups bring unique perspectives - experts providing insights into their fields of knowledge, in this case, the subject of archaeology as well as standards of photographic cataloguing and lay people representing the broader user community that archives and collections aim to serve. By considering both perspectives, this study aims to provide a more nuanced understanding of the potential for generative AI to support practices in archives and collections.

***Research Question 1: Can AI-generated catalogue descriptions be differentiated from human-written descriptions?***

The question of whether machines can generate text that is indistinguishable from human-written text has been a topic of interest since the proposal of the imitation game, often called Turing test (Turing, 1950). The test, which assesses a machine's ability to exhibit intelligent behavior equivalent to, or indistinguishable from, that of a human, has been a benchmark for measuring the performance of artificial intelligence systems in generating human-like texts (e.g. Jakesch et al., 2023; Köbis & Mossink, 2021) and images (e.g. Nightingale & Farid, 2022). In the context of cataloguing, this question takes on a new significance, as the quality of catalogue descriptions are crucial for preserving and providing access to cultural heritage materials.

We thus pose a similar research question whether AI-generated catalogue descriptions can be differentiated from human-written descriptions. If human evaluators cannot distinguish between the two, it suggests that the AI-generated descriptions are similar in quality and usefulness to those written by humans. A distinction between expert and non-expert evaluators is crucial in answering this question. Experts possess a broader knowledge of the images shown and the technical terms used, and a deeper understanding of standards and conventions in cataloguing. This probably allows them to provide a more informed assessment of the AI-generated descriptions. In contrast, non-experts may rely more on general impressions or surface-level features rather than a nuanced understanding of the cataloguing standards. Therefore, we formulated the following hypothesis.

*Hypothesis H1. Experts (vs. non-experts) are better at distinguishing between expert-written and AI-generated descriptions of archival photo material. This holds true irrespective of whether the description was generated by AI or experts.*

Additionally, we measured participants' self-assessed ability to distinguish AI-generated from human-written descriptions as an index of metacognitive calibration, a factor that can sway both trust and willingness to adopt new tools. If participants over- or underestimate this ability, their perception of the model's performance, and thus their readiness to integrate it into cataloguing workflows, may be distorted. Because people often misjudge their own skills and the capabilities of unfamiliar AI systems, we anticipated that self-rated discernment would diverge from actual detection accuracy.

*Hypothesis H2a. Self-assessed ability to distinguish the type of descriptions (self-assessed discernment) measured before the experiment will be significantly different from the actual detection performance (actual discernment) measured at the end of the experiment.*

Further, we hypothesized that experts are worse in their self-assessment because they may be more susceptible to overconfidence due to their extensive knowledge and experience in the field. We tested this assumption in two hypotheses, one regarding the objective classification in expert and non-expert groups and another regarding not only the participants' self-assessed archive expertise, but also their self-assessed expertise in the field of artificial intelligence.

*Hypothesis H2b. The difference between self-assessed discernment vs. actual discernment is expected to be larger for the archive expert group than for the non-expert group.*

*Hypothesis H2c. The difference between self-assessed discernment vs. actual discernment is expected to increase with self-assessed expertise in archives and in AI.*

***Research Question 2: How accurate and useful are the catalogue descriptions perceived to be?***

Second, we assessed the perceived accuracy and usefulness of the AI-generated catalogue descriptions. By comparing the ratings of AI-generated and human-written descriptions, particularly among experts, we can determine whether the AI-generated descriptions are considered accurate and useful. Rating differences between experts and non-experts are important because they reveal how different user groups evaluate the descriptions. Experts with their extensive experience in the field are likely able to differentiate better between useful and accurate information, leading to a less direct correlation between accuracy and usefulness.

*Hypothesis H3. For expert participants, perceived accuracy of the AI-generated descriptions correlates less to perceived usefulness of the AI-generated descriptions compared to non-experts.*

***Research Question 3: How do trust and willingness to use AI tools in general change after seeing the catalogue descriptions?***

Third, we examined the general trust in AI tools and the willingness to use them, which goes beyond the immediate evaluation of the model's performance. Even if the AI-generated descriptions are deemed useful, experts may still be hesitant to adopt the technology due to concerns about trust in AI. We hypothesized that willingness and trust will change because experiencing the AI model's performance firsthand will either alleviate or reinforce existing concerns, leading to an adjustment in participants' attitudes towards AI.

*Hypothesis H4. Participants are expected to adjust their willingness to use AI and trust in AI after distinguishing the descriptions. Pre- and post-measurement of willingness to use AI and trust in AI will be significantly different.*

**Method**

This section describes our work in two parts (overview in Figure 1): Creating the catalogue descriptions as experimental material from LEIZA's image archive materials (Section *Materials*) and conducting the psychological experiment with these catalogue descriptions (Section *Experiment*).

*Figure 1: Materials and experiment*

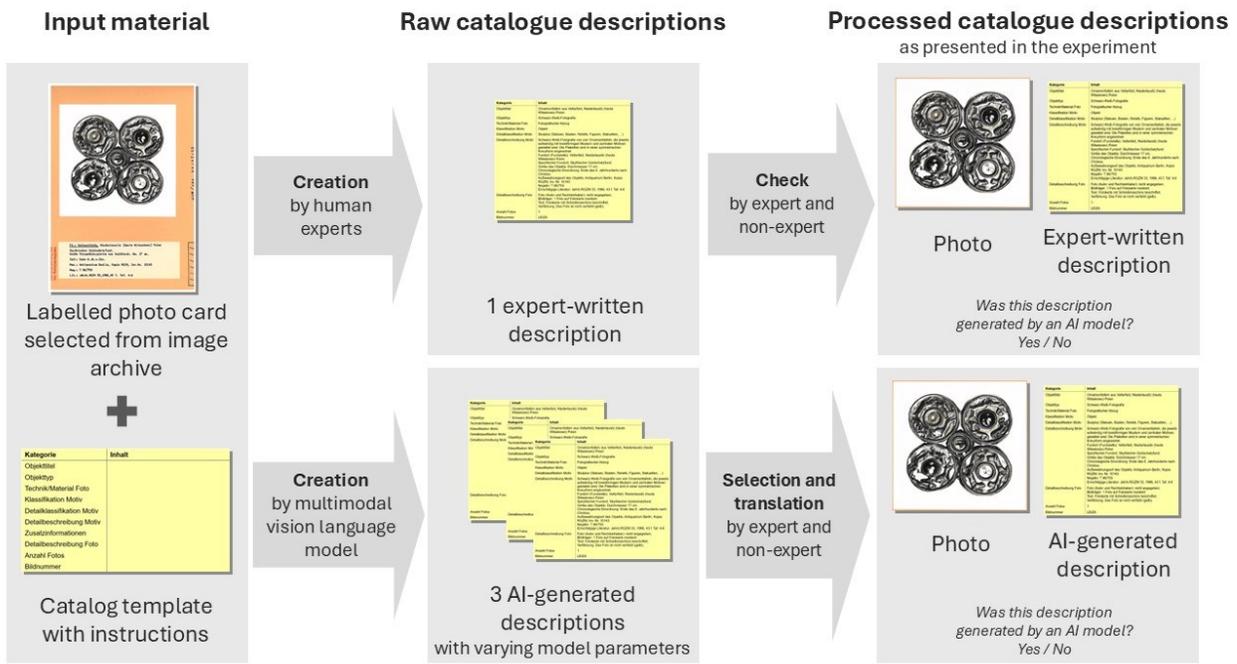

*Note.* Creation procedure of the experimental material and how it was integrated in the experiment.

**Materials**

This subsection describes the creation of experimental materials for which photographic prints mounted on labelled cardboards from LEIZA's image archive functioned as a database. Using a custom-designed catalogue template, human experts and a VLM generated catalogue descriptions from it.

**Labelled photocards as data base for the experimental material**

For the experiment, 40 photographic prints mounted on labelled cardboards (in the following called *labelled photo cards*) were selected from the LEIZA image archive (Klatt, 2021). Each cardboard has one or more photos (of archaeological objects, persons, architecture, landscapes) mounted on it and a hand- or type-written information about the photo (see Figure 2 for an example). Despite the cardboard mounts having different layouts and typing, the semantic structure of the written information on them is comparable. They were designed to make the photos findable to researchers in a twentieth century scientific comparative collection (Bärnighausen et al., 2020). Further, the unique combination of photo and (human-generated) text is particularly beneficial for our study as it allows us to analyze how the vision language model processes visual and linguistic information and combines object recognition and text production tasks.

To be able to generalize as much as possible, we selected labelled photo cards with different dating, layout types, as well as different motive classifications and motive types. To ensure a diverse selection, we included a range of subjects such as archaeological objects, architecture, landscapes, people, and representations from LEIZA's primary main research periods, including Roman times, prehistory, and early medieval times.

*Figure 2: Photographic print on labelled cardboard mount from the LEIZA image archive*

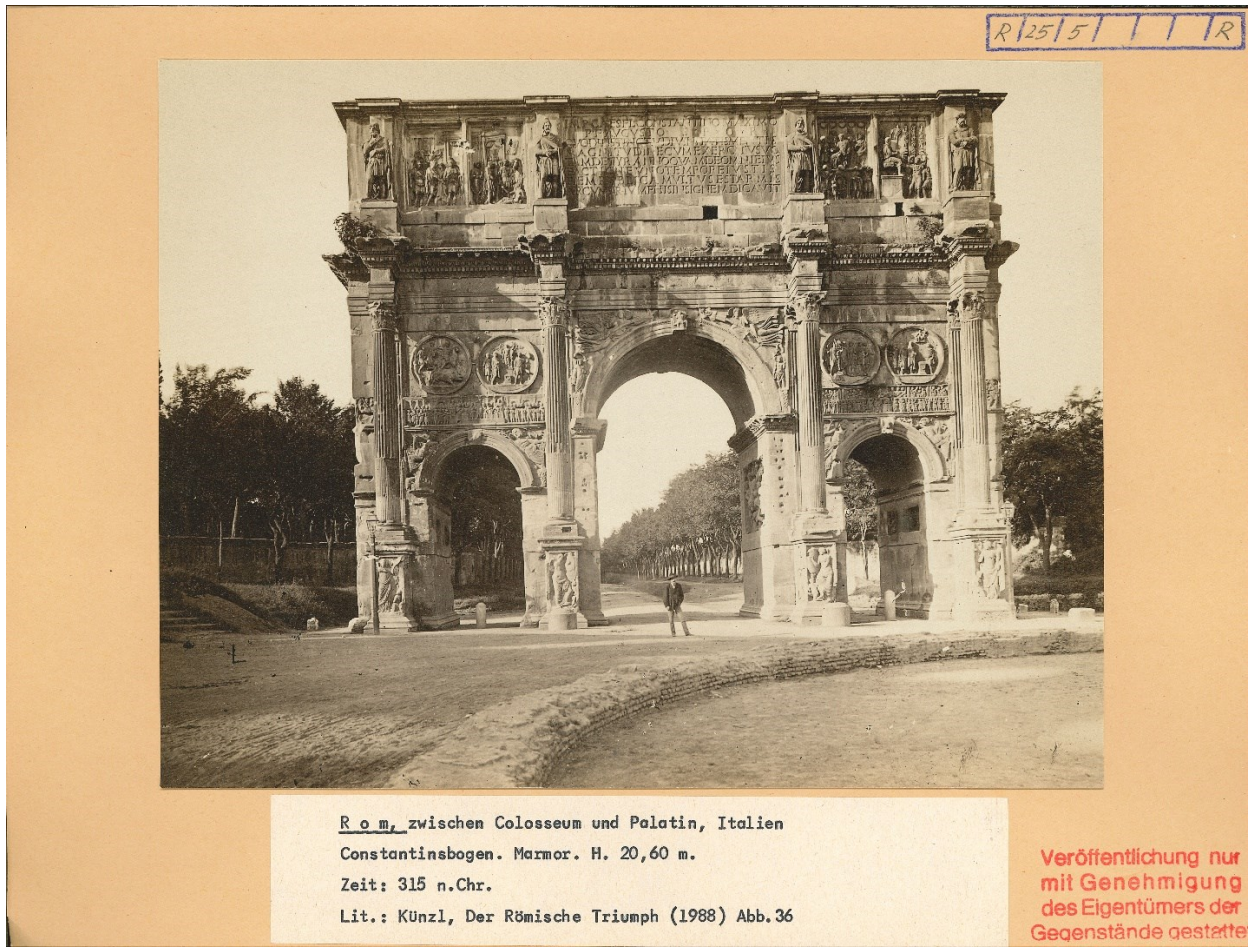

*Note.* Each cardboard has one or more photographic prints mounted on it and a hand- or type-written information text (in the following called *labelled photo card*).

### Designing the catalogue template

To get as close as possible to a realistic recent archiving scenario, we designed a catalogue template to contribute to the aim of creating FAIR digital (meta-)data. In the context of research data management, the "FAIR principles" (https://www.go-fair.org/fair-principles/) enable computational findability, accessibility, interoperability, and reusability (Wilkinson et al., 2016). This meant that the human experts and the VLM were expected to produce comparable descriptions. For this, we developed a catalogue template with data fields (e.g., object title, image number) using the existing historic description categories and the *minimum record recommendation for museums and collections* (*Minimaldatensatz-Empfehlung für Museen und Sammlungen (v1.0.1);* Minimum Record Working Group et al., 2024) as a basis, a broadly supported framework in German GLAM institutions. The minimum dataset is mainly based on the *Lightweight Information Describing Objects (LIDO) Scheme for Metadata of collection objects* by the *ICOM International Committee for Documentation* (*https://cidoc.mini.icom.museum/working-groups/lido/lido-overview/about-lido/what-is-lido/*). To fulfill the experiment's possibilities and reduce complexity, we adapted and reduced the metadata scheme template to ten data fields. For classifying data fields, a selection of terms from accepted controlled vocabularies to classify the objects was suggested in the template. Still, the human experts and the VLM could also use free text. We used controlled vocabulary, for example in the fields "Object type" (*Black and white photography / colour photography / graphic or drawing*) and "Technique or material of the photo" (*Photographic print / print or copy / drawings (original) / paintings (original)*). Categories and vocabulary were iteratively tested with the VLM to ensure they were both useful for archival cataloguing and feasible for AI generation. Instructions for the template (Figure 3) were refined through prompt engineering (Sahoo et al, 2024; Supplement A1.2) and parameter adjustments (Supplement A2).

*Figure 3: Catalogue template*

| Kategorie | Inhalt |
|---|---|
| Objekttitel | |
| Objekttyp | |
| Technik/Material Foto | |
| Klassifikation Motiv | |
| Detailklassifikation Motiv | |
| Detailbeschreibung Motiv | |
| Zusatzinformationen | |
| Detailbeschreibung Foto | |
| Anzahl Fotos | |
| Bildnummer | |

*Note.* This template served as input for both the human experts and the VLM *InternVL2*, to produce comparable catalogue descriptions for each labelled photo card.

**Generating catalogue descriptions**

*Expert-written descriptions.* Expert-written descriptions were created by archive, archaeology, and collections experts from the LEIZA and collaboration partners. The experts specialize in a range of fields relevant to describing or classifying the selected images. This should ensure the necessary competence and a certain level of generalization rather than using just one expert. The experts filled in the template for each labelled photo card following detailed instructions (Supplement A1.1). In plausible cases, they used the existing historic descriptions on the labelled photo cards as a basis, adapted these, and transformed classifications into modern controlled vocabulary. Descriptions were double-checked for typos by an expert and a non-expert but were not edited to create fully homogeneous descriptions.

*AI-generated descriptions.* The open-source VLM *InternVL2-Llama3-76B* (Wang et al., 2024; https://huggingface.co/OpenGVLab/InternVL2-Llama3-76B), processing combined text and image inputs, was selected for its top performance on the *OpenCompass* multimodal leaderboard (as of August 2024;

https://rank.opencompass.org.cn/leaderboard-multimodal), a benchmarking platform that evaluates the performance of multimodal models on image-text tasks by averaging their performance over 8 multimodal benchmarks. The model was not fine-tuned on archival data but instead used with few-shot prompting—instructions that included task explanations and example solutions to guide description generation (Sahoo et al, 2024). The *InternVL2* model received the labelled photo card (image and text) and a detailed prompt (Supplement A1.2) specifying how to fill in the catalogue template. As the main aim of this study is not to evaluate the text extraction capabilities of the model, a human loop was integrated: The model generated three descriptions per labelled photo card with varying parameter values (Supplement A2) from which the non-expert selected the best-fitting description per labelled photo card, ensuring it was free of typos or non-expert detectable errors. It was then translated into German, verified by an expert, and integrated into the experiment.

*Final Materials*

The generation process yielded two descriptions (one expert- and one AI-generated) for each of the 40 labelled photo cards, totaling 80 descriptions (40 expert-written, 40 AI-generated); see Figure 4 for an example pair of descriptions). A combination of a photo (omitting the original labels) with one description constituted an item used in the psychological experiment.

*Figure 4: Example AI-generated and expert-written description*

| Kategorie | Inhalt | Kategorie | Inhalt |
|---|---|---|---|
| Objekttitel | Rom, Konstantinsbogen | Objekttitel | Constantinsbogen, Rom |
| Objekttyp | Schwarz-Weiß-Fotografie | Objekttyp | Schwarz-Weiß-Fotografie |
| Technik/Material Foto | Fotografischer Abzug | Technik/Material Foto | Fotografischer Abzug |
| Klassifikation Motiv | Architektur | Klassifikation Motiv | Architektur |
| Detailklassifikation Motiv | Ehrenbogen oder Triumphbogen | Detailklassifikation Motiv | Triumphbogen |
| Detailbeschreibung Motiv | Schwarz-Weiß-Foto des Konstantinbogens in Rom, Italien. Der Bogen weist drei Hauptbögen auf und ist mit verschiedenen Reliefs und Skulpturen verziert. Der zentrale Torbogen wird von zwei kleineren Torbögen flankiert. Das Bauwerk befindet sich am Ende einer von Bäumen gesäumten Allee. Fundort (FO): Rom, zwischen Kolosseum und Palatin; zeitliche Einordnung (Zeit): 315 n. Chr. (315 AD); relevante Literatur: Künzl, Der Römische Triumph (1988) Abb. 36; | Detailbeschreibung Motiv | Schwarz-Weiß-Ansicht des Constantinsbogens in Rom, zwischen Colosseum und Palatin; In der Bildmitte posiert ein Mann mit Hut und Bekleidung, die auf das frühere 20. Jh. hinweist; Ortsbezug: Rom, Italien; zeitliche Einordnung: römische Kaiserzeit, um 315 n. Chr.; Literaturangabe: E. Künzl, Der Römische Triumph, 1988, Abb. 26; |
| Detailbeschreibung Foto | Fotograf und Herkunft des Fotos: nicht angegeben | Detailbeschreibung Foto | Foto (Urheber und Rechteinhaber): unbekannt, ev. Ernst Künzl?; Foto gelb-bräunlich verfärbt; Bildträger: 1 Foto auf Bildkarte montiert; Beschriftung: Karte mit Schreibmaschine beschriftet; |
| Anzahl Fotos | 1 | Anzahl Fotos | 1 |
| Bildnummer | Nicht angegeben | Bildnummer | |

*Note.* Example catalogue descriptions, one generated by the VLM (left) and one written by a human expert (right).

**Experiment**

The experiment was preregistered on the Open Science Framework (https://osf.io/cy4hg/?view_only=fe01f024ce6340ee94f91f528c0002be). All presented question items can be found in Supplement A3.

*Experimental setup and procedure*

The experiment was programmed in PsychoPy (Peirce et al., 2019) and translated into JavaScript code to make it accessible as an online experiment via Pavlovia (https://pavlovia.org/). It consisted of three phases. In the first phase, participants were asked about their willingness to use AI and their trust in AI (on a seven-point Likert scale). An example trial showed the photo from the labelled photo card with one of the corresponding descriptions and asked about the description's accuracy, usefulness, and whether an AI created it. Participants then estimated their classification performance for the following trials, that is, they estimated how many of the following 80 items they would classify correctly as AI-generated or not.

The second phase presented the 80 items (40 photo-text combinations with expert-written descriptions, 40 photo-text combinations with AI-generated descriptions) to the participants, distributed over two blocks with 40 trials each, ensuring that all participants were confronted with all items during the experiment. We counterbalanced whether participants saw the AI-generated or the expert-written description along with the corresponding picture in the first or the second block. Inside each block, the items were presented in a random order.

In each trial, participants responded to four questions for the presented item: They first had to rate the accuracy and usefulness of the description. Then, they were asked whether the description was generated by an AI model or not. Finally, participants had the opportunity to add comments on the description in a free text field before the following trial started with a new image and a corresponding description. After 40 trials, a break screen was presented before participants continued in the second block with the remaining 40 descriptions. After the second block, participants were provided with their actual detection performance.

Phase 3 repeated the questions on willingness to use AI and trust in AI. Participants were also asked to rate their expertise in the field of archival sciences, archaeology, and AI (all on a seven-point Likert scale). Furthermore, we assessed different fields of expertise as a categorical variable by asking whether participants work or study in the field of archival sciences and/or AI. At the end of the experiment, participants had the opportunity to withdraw their consent to use their data for research purposes before they were led to a separate page where their data for compensation (payment or course credits) was collected.

*Experimental design*

We manipulated the factors *description type* (AI-generated and expert-written descriptions; within-subjects) and *expertise* (expert and non-expert; between-subjects) in a 2x2 mixed design. The factor *expertise* was measured in two ways: We asked for participants' jobs and study programs (expert vs. non-expert group, categorical with several options to choose which were aggregated into an expert and a non-expert group) and for a self-assessment of their expertise in archives, archaeology as well as in AI (seven-point Likert scale, used in H2c). For a better understanding of our experimental data's composition and exploratory analyses, we additionally assessed AI expertise as a categorical variable (expert and non-expert).

Dependent variables were the *classification performance* for AI-generated and expert-written descriptions measured in binary responses per trial and later transformed into measures of Signal Detection Theory (Green & Swets, 1966), the *difference between self-assessed and actual classification performance* (in %), *perceived accuracy* of the description from 0 to 100%, *perceived usefulness* of the description from 0 to 100%, *willingness to use AI* (two items), and *trust in AI* (one item), both used as pre-test scores (measured in Phase 1), post-test scores (measured in Phase 3), and differences in pre- and post-test scores.

*Data collection*

Data collection went from the beginning of December 2024 to the end of January 2025. We recruited participants via newsletters at the University of Tübingen, personal e-mails to archivists, printed flyers distributed at the University of Tübingen, and announcements at the archival conference "Artificial Intelligence in Archives and Collections" in December 2024 in Marburg, Germany.

## Experimental evaluation

**Employed analyses**

We pre-processed the experimental data and analyzed it with the pre-registered frequentist method and additional Bayesian statistics for a deeper understanding of the results. All data and analysis scripts can be downloaded from *zenodo[i]*.

*Data pre-processing*

The raw experimental data was pre-processed in several steps. Categorical expertise responses were aggregated into *expert* (including all archaeology and archive experts, see Supplement A3) and *non-expert* groups. For H1 on classification performance, we calculated measures from Signal Detection Theory (SDT; Green & Swets, 1966) using the *psycho* package (Mackowski, 2018). This mathematical framework measures the ability to differentiate between two stimuli, called signal (bearing information) and noise (distracting from the signal). In our study, we treat AI-generated descriptions as the "signal", whereas the expert-written descriptions are the background "noise". For each participant, we calculated d', the SDT measure of sensitivity, telling us how accurately they can differentiate an AI-generated description from an expert-written one (a d' of 0 means they are guessing, i.e., chance-level performance). To see whether participants tend to prefer one response over the other, we also computed SDT's c (response bias): a positive c indicates a general inclination to label descriptions as expert-written, whereas a negative c indicates an inclination to label them as AI-generated.

*Frequentist analysis*

Frequentist analysis is a statistical approach that focuses on the frequency of events, allowing us to determine whether our results are due to chance or reflect a real pattern in the data. For the preregistered analysis, we checked the normal distribution of the data and used t-tests (with significance level $α$ = .05) with Holm correction (Holm, 1979) when doing multiple comparisons, and comparisons of correlation tests (*cocor* package; Diedenhofen & Musch, 2015) for all four hypotheses to examine whether the analyzed data provides strong enough evidence to suggest that the null hypothesis (describing no effect) is unlikely to be true. Additionally, we fitted generalized linear mixed-effect models (GLMMs) to predict one measured response variable from other variables, called fixed and random factors. We used GLMMs of the logit-family (*lme4*-package; Bates et al., 2015) to predict participants' binary classification performance (correct/incorrect) from *description type* and *expertise* to investigate H1 further. To double-check our main findings, we extend the models with the factors *self-assessed classification performance*, *trial number*, *accuracy ratings,* and *usefulness ratings* in exploratory analyses to analyze their influence on classification performance.

### *Bayesian analysis*

To complement the frequentist analyses, we used Bayesian statistics, which enables explicit testing of the null hypothesis. We computed Bayesian t-tests that provide a Bayesian alternative to traditional frequentist t-tests, estimating the probability of a difference between two hypotheses and quantifying the magnitude of this difference as a Bayes factor (BF). Following Jeffreys (1961), BFs above 1 provide evidence in favor of the alternative hypothesis (1-3 anecdotal, 3-10 substantial, 10-30 strong, 30-100 very strong, and > 100 extreme evidence), while BFs below 1 represent evidence for the null hypothesis.

**Results**

*Participants*

After excluding participants who did not consent to data usage (*N* = 1) or who did not complete the online experiment (*N* = 60), the final sample consisted of *N* = 139 participants, including 110 non-experts and 29 archive and archaeology experts, aged between 18 and 67 years (*M* = 25.20, *SD* = 8.14), with one non-binary, 43 male and 95 female participants.

*Hypotheses*

**Hypothesis 1: Classification performance by expertise**

To test for an expert advantage in the classification of expert-written and AI-generated descriptions (H1), we began with a signal-detection analysis. Treating AI-generated descriptions as the "signal" and expert-written descriptions as the "noise," we calculated sensitivity (d') for each participant. Mean d' was reliably above zero (*M* = 0.67, *SD* = 0.62), *t*(138) = 12.67, *p* < .001, indicating that, overall, people could tell machine- from human-written text. Contrary to our prediction, experts (*M* = 0.83, *SD* = 0.76) did not outperform non-experts (M = 0.62, SD = 0.57), *t*(36.91) = 1.37, *p* = .180.

A Bayesian check confirmed these conclusions. The Bayes factor for the overall d' (*BF* = 5.6 × 10$^{21}$) provided extreme evidence for genuine discriminability, whereas the factor contrasting experts and non-experts (*BF* = 0.68) offered only moderate evidence for the null, again suggesting no expertise advantage.

Because every trial required the same binary decision ("Was this description produced by AI?") the d' metric already captures the distinction between description types, making the preregistered ANOVA superfluous. Instead, we modelled trial-by-trial accuracy with a generalized linear mixed model (GLMM). The model included description type and expertise as fixed effects and allowed the effect of description type to vary by participant to absorb individual response biases (i.e., random slopes). A type

II ANOVA on this fitted model revealed a main effect of description type, $\chi^2(1)$ = 31.58, $p$ < .001: participants more often classified expert-written descriptions correctly than AI-generated ones. Neither the main effect of expertise, $\chi^2(1)$ = 2.70, $p$ = .101, nor the interaction, $\chi^2(1)$ = 1.13, $p$ = .290, reached significance. Accuracy therefore ranged from 54.9 % (non-experts on AI texts) to 67.8 % (experts on expert texts; see Figure 5 and Supplementary Table B1.1).

Taken together, participants could distinguish AI-generated from human-written catalogue entries at a moderate level, and this ability did not depend on archival or archaeological expertise; nonetheless, human texts seem to be identifiable more easily than machine texts, although response biases towards the human texts may have contributed to this effect (see exploratory analysis of response bias).

*Figure 5: Classification performance by expertise and description type*

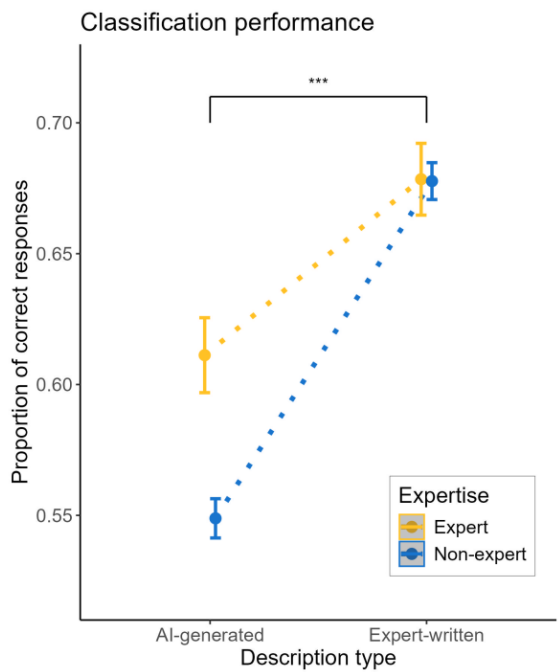

*Note.* Error bars indicate the standard error of the mean (SEM).

**Hypotheses 2: Self-assessment of classification performance**

To evaluate metacognitive calibration (H2a–H2c), we compared participants' self-rated ability to spot AI descriptions with their actual detection accuracy. Mean self-estimates ($M$ = 49.23 %, $SD$ = 15.33) fell well below true performance ($M$ = 61.99 %, $SD$ = 10.72), $t(246.88)$ = 8.04, $p$ < .001, confirming the difference expected in H2a: participants' initial expectations about their performance were more pessimistic than their actual performance warranted. The size of this gap did not differ by professional status—experts ($M$ = –15.44 %, $SD$ = 18.52) were no less calibrated than non-experts ($M$ = –12.06 %, $SD$ = 18.44), $t(43.79)$ = –0.88, $p$ = .386; the corresponding Bayes factor ($BF$ = 0.31) offered no substantive evidence for or against an expertise effect. Nor did the discrepancy correlate with participants' self-reported expertise in archives, archaeology, or AI (Supplementary Table B1.2), countering H2c. Finally, adding self-assessment scores as a predictor in the GLMM left classification accuracy unchanged, $\chi^2(1)$ = 0.21, $p$ = .646. In short, participants generally low-balled their own detection ability, and this miscalibration was unrelated to domain expertise.

**Hypothesis 3: Perceived accuracy and usefulness of the descriptions**

To probe H3, we asked whether accuracy and usefulness are weighted differently by experts and laypersons when they judge the AI output. For AI-generated descriptions, experts' ratings of the two dimensions were almost perfectly aligned, $r(27)$ =.87, $p$ <.001, whereas non-experts showed a weaker association, $r(108)$ =.69, $p$ < .001 (see Figure 6). A Fisher z-test confirmed that the expert correlation was significantly stronger, $z$ = 2.20, $p$ = .014 (an effect that replicated for expert-written texts, see Supplementary Table B1.3). Thus, contrary to H3, experts treated factual accuracy and practical usefulness as more, not less, intertwined than non-experts did.

Mean ratings told a parallel story: across the full sample, human descriptions were judged both more accurate, $t(275.36)$ = -5.62, $p$ < .001, and more useful, $t(275.77)$ = -6.88, $p$ < .001, than their AI counterparts (see Supplementary Table B1.4 and Figure 6). Together, these results show that AI-generated catalogue entries still lag behind human prose on core quality metrics and that experts integrate those metrics more tightly when forming their evaluations.

*Figure 6: Accuracy and usefulness ratings of expert-written and AI-generated descriptions by expertise*

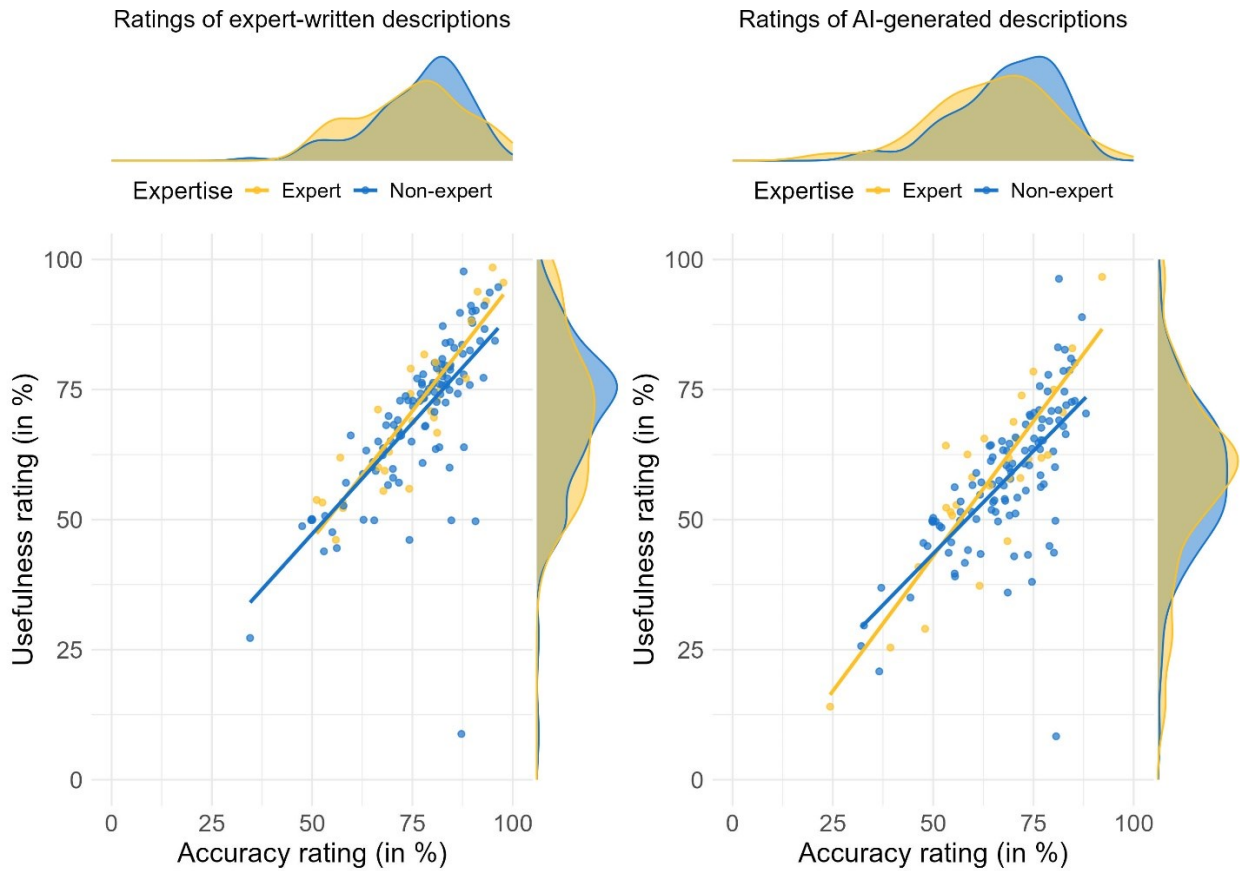

*Note*. Lines indicate the correlations between accuracy and usefulness ratings. Distributions of ratings are shown in the areas above and on the right of the plots.

**Hypothesis 4: Willingness to use and trust in AI**

To test H4, we compared willingness-to-use and trust scores before and after the classification task. Participants entered the study modestly positive about AI tools in general (willingness: *M* = 5.43, *SD* = 1.63; trust: *M* = 3.86, *SD* = 1.26) but left noticeably less enthusiastic (willingness: *M* = 5.09, *SD* = 1.65; trust: *M* = 3.66, *SD* = 1.40). The declines were reliable—willingness, *t*(138) = 4.33, $p_{Holm}$ <. 001; trust, *t*(138) = 3.11, $p_{Holm}$ = .005—indicating that direct exposure to the model's mixed performance tempered respondents' general readiness to adopt and their confidence in AI tools (see Figure 7). In short, hands-on evaluation reinforced existing reservations rather than alleviating them.

*Figure 7: Willingness to use and trust in AI models*

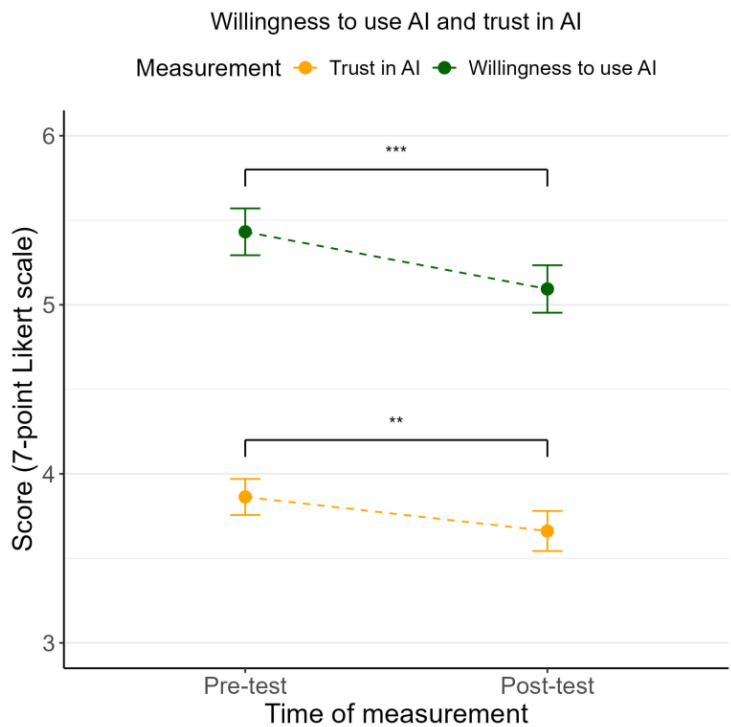

*Note*. Error bars indicate standard errors of the mean (SEM). The corresponding questions posed to the participants focused on AI tools in general rather than the specific model used in the study.

*Exploratory analyses*

To probe the boundary conditions of our preregistered results, that is, to test how far the main effects hold once we vary task dynamics and individual attitudes, we ran four supplementary analyses. Specifically, we examined (1) whether classification accuracy changed as participants gained experience across trials, (2) whether systematic response biases emerged, (3) how perceived accuracy and usefulness of AI text affected detectability, and (4) whether expertise shaped post-task willingness to adopt and trust AI tools. Together, these probes reveal how learning, decision heuristics, content quality, and professional identity modulate the headline effects.

### Classification performance over time

To check whether participants familiarized themselves with the task, we assessed whether their classification performance evolved over time by fitting an exploratory GLMM including *trial number* as a fixed effect. This analysis confirmed a significant improvement in performance as participants progressed through the task, with the proportion of correct responses increasing significantly with trial number, $\chi^2(1)$ = 4.83, *p* = .028. This learning effect suggests that participants became more adept at distinguishing expert-written and AI-generated descriptions over time.

### Response bias

To observe any response classification tendencies participants might have for certain response categories and to double-check our main finding of participants' higher classification performance of expert-written descriptions, we additionally computed SDT's response bias *c* of the participants. Results revealed a positive mean bias (*c* = 0.18), significantly different from zero, *t*(138) = 5.39, *p* < .001, indicating a systematic tendency to favor "expert-written" classifications—a finding consistent with and probably contributing to the GLMM's main effect, which showed higher performance in classifying expert-written descriptions. Notably, no significant differences emerged in response bias between experts (*M* = 0.11, *SD* = 0.32) and non-experts (*M* = 0.20, *SD* = 0.41), *t*(55.34) = 1.24, *p* = .219, suggesting both groups applied similar decision criteria despite differences in expertise.

**Influence of accuracy and usefulness ratings on classification performance**

For validating our Turing test approach in research question 1 (indistinguishable AI and human texts suggest that AI texts are similar in quality), we reverted the approach and examined how participants' perceived accuracy and usefulness of the AI-generated descriptions might influence classification performance. For that, we fitted two GLMMs with *accuracy ratings* and *usefulness ratings* as an additional fixed factor. Both analyses revealed significant main effects: Higher perceived accuracy of AI-generated descriptions was associated with reduced classification performance, $\chi^2(1)$ = 50.75, $p$ < .001, and higher perceived usefulness similarly predicted lower classification performance, $\chi^2(1)$ = 64.40, *p* < .001. This can be explained by AI-generated descriptions appearing more similar to expert-written descriptions with increasing accuracy and usefulness, and thus becoming harder to classify as AI-generated. These results provide evidence that classification performance is sensitive to the perceived quality of AI-generated content, thereby validating our approach in research question 1 to operationalize classification performance as a proxy for the human-likeness of AI-generated descriptions.

**Expert-specific willingness to use AI and trust in AI**

To assess how expertise influences the adoption of AI tools in archival practices, we conducted an exploratory multivariate ANOVA (all four pre- and post-measures of willingness and trust ~ expertise*)* followed by four separate univariate ANOVAs with Holm correction (Holm, 1979)*.* Reported willingness to use AI was lower for experts than for non-experts (exact values in Supplementary Table B1.5), both for measures before, $F_{pre}(1, 137)$ = 8.75, $p_{Holm}$ = .015, and after the test, $F_{post}(1, 137)$ = 6.43, $p_{Holm}$ = .037.

Regarding trust, there was no difference in pre-trust scores, $F_{pre}(1, 137) = 3.41$, $p_{Holm} = .134$, and no difference in post-trust scores between experts and non-experts, $F_{post}(1, 137) = 1.16$, $p_{Holm} = .284$.

Together, these exploratory analyses show that participants learn quickly, carry a modest "expert-written" bias, find high-quality AI descriptions hardest to detect, and become more cautious about adopting AI tools after hands-on evaluation.

## Discussion

This study examined the potential of generative AI, specifically a VLM, to be meaningfully integrated into the cataloguing of photographic materials in archives and collections. Central to our *human-centered* investigation was the question of whether AI-generated catalogue descriptions can match or approximate the accuracy, usefulness, and perceived trustworthiness of those written by professional collection managers or archivists. To answer this question, three research questions were posed and evaluated using an *experimental framework* in which archive and archaeology experts, as well as non-experts, classified and rated AI-generated and expert-written catalogue descriptions and reported willingness to use and trust in AI tools. Both, experts and non-experts, were able to classify them as AI- and human-generated. They perceived expert-written descriptions as more accurate and useful than AI-generated descriptions and showed lower willingness to use and trust in AI tools after the classification task.

**Research Question 1: Can AI-generated catalogue descriptions be differentiated from human-written descriptions?**

Our idea, inspired by the Turing test, was that if human evaluators could not distinguish between the two descriptions, it suggests that AI-generated descriptions are similar in quality to those written by humans. This is supported by our result that AI-generated descriptions rated higher in perceived accuracy and usefulness became harder to classify. Our key finding shows that participants were able to differentiate between AI-generated and expert-written descriptions, even the non-experts. This means that the out-of-the-box VLM, which was not trained on domain-specific datasets like the labelled photo cards, failed to pass the Turing test. As a result, it may not yet be suitable for generating high-quality texts on image-text sources according to cataloguing standards fully on its own, highlighting the need for human oversight and review.

Further, participants reported lower self-assessed classification performance before the task than their actual performance warranted. This discrepancy could be due to high expectations towards the AI texts before the classification task (and thus a low estimation of one's own discernment ability), which were then not met by the actual experience. While expertise in different fields (archive, archaeology, AI) did not influence this self-assessment, it may also reflect a general unfamiliarity with AI models and their current capabilities, which could pose a barrier to meaningful integration in cataloguing processes.

**Research Question 2: How accurate and useful are the catalogue descriptions perceived to be?**

Expert-written descriptions were rated as more accurate and useful than AI-generated ones. Further, accuracy and usefulness ratings were more strongly correlated among experts than non-experts, suggesting that experts' perceptions of accuracy were more tightly linked to their perceptions of usefulness. This finding contrasts with our initial hypothesis, which assumed that experts, due to their domain-specific knowledge, would exhibit a clearer distinction between these dimensions. However, the found pattern may reflect the practical integration of accuracy and usefulness as complementary criteria in cataloguing: For professionals, ensuring factual precision (accuracy) and functional applicability (usefulness) could often be inseparable priorities in creating descriptions.

**Research Question 3: How do trust and willingness to use AI tools in general change after seeing the catalogue descriptions?**

Our findings reveal a post-exposure decline in trust and willingness to use AI tools, consistent with prior work showing that technical performance alone is insufficient for AI adoption in archival contexts (Jaillant & Rees, 2023). This decline may stem from concerns about the AI's lack of transparency, as participants, particularly professionals attuned to metadata risks (Bunn, 2020; Gilliland, 2012), may perceive unclear decision-making as amplifying errors. A more transparent setup, such as presenting alignment with applied standards during material creation to participants, could mitigate skepticism (Yu & Li, 2022). Notably, while willingness to use AI remained slightly higher than trust scores, this discrepancy suggests a tension between curiosity and caution: participants expressed interest in exploring AI's potential but remained wary of its risks. Together with experts' consistently lower willingness scores, this underscores the need to systematically evaluate professionals' acceptance and perceived risks of AI tools and to actively engage their concerns, particularly their demand for transparency and alignment with ethical stewardship, to advance the archival debate from "can the system perform?" to "on what terms will professionals accept its results?" (Davet et al., 2023).

Taken together, our results show that performance metrics alone cannot secure adoption: Participants, especially experts, preferred human authorship, and their willingness to use AI fell after hands-on experience. Archivists' and collection managers' orientation toward provenance and evidential integrity magnifies even minor hallucinations; when outputs gain fluency without exposing their decision trail, uncertainty deepens. This echoes algorithm-aversion studies in which users avoid relying on algorithms after seeing none of them err (Dietvorst et al., 2015). Human-centered human-computer interaction research demonstrates that participatory methods—scenario co-creation, think-aloud walkthroughs, question-driven explainability—surface such biases and calibrate trust (Liao et al., 2021; Hashmati et al., 2024). GLAM case work confirms that co-design with curators improves acceptance by embedding AI within existing governance routines (Co et al., 2023; Digital Scholarship Group, 2020). These strands converge with the "participatory turn" in AI design, which argues that domain stakeholders must shape requirements from the outset (Delgado et al., 2023).

Accordingly, a cataloguing AI pipeline should encode provenance and accountability as first-order constraints, involve archivists and collection managers throughout development, and make every machine recommendation transparent, auditable, and reversible (Jaillant & Rees, 2023; Shneiderman, 2022). Aligning system affordances with professional cognition and ethics is therefore not an optional add-on but the decisive factor in turning generative models from promising prototypes into trusted collaborators (Davet et al., 2023). To achieve this, future implementations must address two key aspects: technical advancements and trust-building measures.

**Future directions**

These two aspects are intertwined, as improving model performance through domain-specific fine-tuning on archival data can reduce hallucinations and OCR inaccuracies, while building trust requires training archivists and collection managers to critically engage with AI and balance efficiency gains with curatorial responsibility for preservation. By combining these approaches, we sketch the idea of an explainable AI pipeline that provides transparency into the output generation process, ultimately leading to more reliable and trustworthy AI systems.

*Technical advancements.* Despite prompt engineering and model parameter adjustments, small OCR errors and hallucinations in AI-generated descriptions persisted. This highlights the current limitations of off-the-shelf models in cataloguing contexts, at least in multimodal applications, and underscores the need for domain-specific fine-tuning to achieve human-like quality. Fine-tuning involves adjusting the parameters of a pre-trained AI model to fit the specific characteristics and nuances of a particular domain, in this case, archaeological material. This process involves training the model on a large dataset that is representative of the specific domain, such as the labelled photo cards used in this study. By doing so, the model can learn to recognize and replicate the specific types of objects, people, and events depicted in the images, and can combine this with the unique linguistic and stylistic features of written text. This can reduce the occurrence of hallucinations and OCR inaccuracies, leading to more reliable and accurate descriptions.

*Trust-building measures*. The successful integration of AI models into cataloguing depends on more than just technical improvements. It requires a collaborative approach that balances the benefits of AI with the need for human oversight and expertise. Our findings advocate for a collaborative model where AI serves as a drafting tool for initial description generation, while archive and archaeology experts retain stewardship roles in verification, oversight, and contextual refinement. This hybrid approach aligns with risk-opportunity perception research (Schwesig et al., 2023) as well as ethics frameworks emphasizing that AI integration must not compromise transparency, provenance, or negotiable criteria (Jaillant & Rees, 2023). Future implementations should thus focus on building trust. This could, for example, be achieved by the provision of targeted training and education for archaeology and archive experts to critically engage with AI tools and to balance efficiency gains with curatorial responsibility for preservation. For example, a training program might include hands-on exercises where experts work with AI tools to generate descriptions of sample images or texts, and then evaluate and refine the outputs based on their own expertise and knowledge. This would help experts to develop the skills and confidence they need to work effectively with AI tools, while also ensuring that the outputs are accurate, reliable, and consistent with standards in GLAM institutions.

*Combining technical advancements and trust-building measures.* A concrete vision combining both model improvements and trust-building measures is the development of an explainable AI pipeline that provides transparent, auditable, interpretable, and reversible outputs, allowing users to understand how the AI model arrived at its conclusions. For example, an explainable AI pipeline for image description generation might provide a visualization of the image features that were most relevant to the generation of the description, such as specific objects, colors, or textures. This would allow researchers and the public alike to understand how the AI model interpreted the image and to verify that the description is accurate and relevant, leading to more reliance on the model and the human-AI collaboration.

**Conclusion**

Our study demonstrates that catalogue descriptions generated by a not fine-tuned model can approach, but not yet fully match, the quality of human-written descriptions. However, with domain-specific fine-tuning and ongoing advancements in VLMs, models can or will shortly be able to produce catalogue descriptions that appear human-like to lay and professional evaluators. Further, our results show that performance metrics alone cannot secure successful adoption, but that establishing trust is essential. We therefore advocate for an explainable AI pipeline providing transparent, auditable, interpretable, and reversible outputs which, in turn, enables GLAM institutions to harness the benefits of AI models while ensuring that the integrity and accuracy of their collections are maintained.

## Data availability

All materials, code, and datasets generated and analyzed during the study are available in a *zenodo* record[ii].

## Supplementary information

Supplementary materials contain information on material creation and additional results in tables.

## Acknowledgements

We want to thank all of Dominik Kimmel's colleagues for creating the descriptions.

## Ethics declarations

**Ethics Approval**

This study was performed in line with the principles of the Declaration of Helsinki. Approval was granted by the local ethics committee of the Leibniz-Institut für Wissensmedien (Date: October 2, 2024; No: LEK 2024/036).

**Informed consent**

Experimenters obtained written consent from the participants at the beginning (on participation and data use) and at the end of the online experiment (data use and consent to publish).

**Competing interests**

The authors declare no competing interests.

**AI disclosure**

A VLM was used for material creation as explained in the Methods section. Large language models (*ChatGPT, LLaMA, Qwen*) were used for language polishing in manuscript preparation.

## Author contributions

**Authors and Affiliations**

This can be found on the title page.

**Contributions**

All authors contributed to this research and manuscript. Design of material: first, second, second to last and last author. Research questions and hypotheses: all authors. Experiment set-up: first and third authors. Data analysis: first author, third author, last author. Manuscript writing: First author, second to last, last author. Critical feedback and editing of manuscript: All authors.

**Corresponding author**

Correspondence to: Markus Huff, m.huff@iwm-tuebingen.de

## Supplementary information

This file includes supporting information for the manuscript "ArchiveGPT: A human-centered evaluation of using a vision language model for image cataloguing", structured as follows:

Supplementary information A: Materials and experiment

    A1. Instructions and prompts

        Supplements A2.1 Instructions used for experts

        Supplements A2.2 Engineered prompt used for the model

    A2. Model parameters

    A3. Items presented in the experiment

Supplementary information B. Results

    B1. Supplementary tables

        Table B1.1 Predicted classification performance

        Table B1.2 Correlations of self-assessed expertise with self-assessed classification performance

        Table B1.3 Correlation of accuracy and usefulness ratings and their comparisons

        Table B1.4 Accuracy and usefulness ratings

        Table B1.5 Expert-specific willingness to use AI and trust in AI

# Supplementary Information A. Materials and experiment

## *Supplementary Information A1. Instructions and prompts*

### Supplementary Information A1.1: Instructions used for experts

**Beschreibung der Bildkarten: Prototyp**

Zusammenfassend nach Vorbild Minimaldatensatz/Vokabulare: AAT soweit möglich

| Feld | Beispielbezeichnungen, Auswahlkategorien, | | Zuordnung |
|---|---|---|---|
| | | | |
| **Objekttitel:** | | | Motiv |
| **Objekttyp:** | <ul><li>Schwarz-Weiß-Fotografie</li><li>Farb-Fotografie</li><li>Grafik</li><li>Gemälde</li></ul> *(<< es muss einer Kategorie zugeordnet werden* | | Foto |
| **Technik/Material (Bildobjekt)** | <ul><li>Fotografischer Abzug</li><li>Druck</li><li>Zeichnung</li><li>Gemälde</li></ul> *(falls bekannt ergänze jeweils Detailbegriff. >> Excel Liste)* | | Foto |
| **Klassifikation Motiv**: | <ul><li>Objekt (object) (inkludiert insb. Archäologisches Objekt (archaeological object, cultural artifact)</li><li>Siedlungen und Landschaften (Settlements and Landscapes)</li><li>Architektur (architecture)</li><li>Personen (people)</li></ul> *(<< es muss einer Kategorie zugeordnet werden)* | | Motiv |
| **Detailklassifikation Motiv** | s. Liste Excel | | |
| **Detailbeschreibung**: | Schwarz-Weiß-Fotografie eines …; *(kurze Beschreibung des sichtbaren Motivs, soweit möglich Material, Technik und Funktion beschreiben)*; Text gerne etwas länger und beschreibend; Fundort: … ; zeitliche Einordnung: … ; Datierung: … ; Verwahrort d. Objekts: … ; <br><br>(*bei Personen*: „Angaben zur Person"; *bei Architektur u Landschaft*: „Ortsbezug"; <br><br>*(wenn nichts bekannt ist, bleiben die Angaben leer)* | | Wichtigstes prägnant ; Verbindung aus Foto u Motiv |
| **Anzahl Fotos:** | 1 | | Foto |
| **Details Foto** | Foto (Urheber und Rechteinhaber): (*soweit erkennbar oder bekannt*) ……. ; <br>Bildträger: 1 (2, …) Fotos auf Bildkarte montiert; <br>Beschriftung: Karte mit Schreibmaschine beschriftet / Karte handschriftlich beschriftet); | | Fotoobjekt |
| **Bildnummer** | Eindeutige Nummer des Fotos (Neg.) so wie am Objekt beschrieben. | | |
| | | | |

**Kommentare zum Ausfüllen für Experten und Prompts:**

Beschreibungen/Angaben sollten so gemacht werden, wie sie üblicherweise als Experte/ Archäologe beim ausführlichen katalogisieren gemacht werden und dürfen wahlweise kurz, knapp, stichwortartig sein oder aber in Sätzen. Idealerweise sollten sie etwas länger sein als die unten genannten Beispiele und über den vorliegenden Text auf der Bildkartei hinausgehen. Wenn möglich auch noch eine kurze Beschreibung wichtiger Merkmale ergänzen, auch wenn diese nicht auf dem Text der Bildkarte vermerkt sind. Für dieses Experiment bitte auch - wenn sinnvoll - auch kurze Beschreibungen zum Sichtbaren ergänzen (< das kann eine künftige Volltextsuche /bzw. Erschließung nach Keyword erleichtern oder z.B. Blinden das Nutzen ermöglichen). Für das Experiment soll/kann zusätzliches Expertenwissen in die Beschreibungen einfließen aber eher keine längeren zusätzlichen Infos aus der Literatur etc.. Verweise auf Literatur sind möglich aber nicht nötig. (Wir testen an dieser Stelle nicht in erster Linie die Fähigkeit von ChatGPT Datenbanken zu durchsuchen). Kleinere Präzisierungen und Erläuterungen, die für das Verständnis sinnvoll sind, sind ok.

Wenn erkennbar, dass es sich um ein Reprophoto aus anderer Quelle handelt, kann das gerne beschrieben werden.

**Zu Feld Details:**

Das Feld ist eine Zusammenfassung der wichtigsten Infos zum Motiv (arch Objekt, Person etc.) sowie zum Fotoobjekt. Es soll prägnant beschrieben werden, so dass es auch für eine allfällige Übergabe an digitale Bibliotheken (z.B. DDB) geeignet ist.

- Soweit möglich bei Orten (Fundorten, Verwahrorten (Museen), Fundstätten, etc. Stadt und Staat dazu schreiben in moderner heutiger Schreibweise (<< als Vokabular wird dann GDN verwendet).
- In der Beschreibung i.R. immer die Infos zum Motiv direkt als erstes zusammenfassen soweit möglich nach dem Vorbild; stichwortartig oder kurze Sätze; ebenso zusammenfassen Details zum Fotoobjekt (zb. Material, Provenienz und Urheberschaft. (<< siehe auch Beispiele)
- im Feld „zeitliche Einordnung" nach DAI Chrontology ChronOntology (dainst.org) (s. auch Liste: << die zeitliche Einordnung spiel im Test aber keine wichtige Rolle, da ChatGPT hier wohl nur das erkennt was es im Text lesen kann >> daher hier nur grob einordnen (römisch, römische Kaiserzeit, Hallstattzeitlich, …)

**Zum Feld Details Fotoobjekt:**

Hier können alle Infos zum Fotoobjekt erfasst werden.

- Bedingungen / Erklärungen zum Fotourheber
- wenn ein Museum o.ae bei der Bildnummer (Neg. Nr) dabei steht, dann ist das I.R. ein Foto, das nicht vom LEIZA stammt.
- Wenn in Neg. eine Nummer ohne weitere Ortsangabe steht, dann i.R. ein LEIZA
- wenn Feld auf Karton leer: „nicht bekannt"

**Beispiele:**

# ArchiveGPT 43

Kurze Version, ohne viel Beschreibung;

| Feld | Beispielbezeichnungen, Auswahlkategorien, | | Zuordnung |
|---|---|---|---|
| **BK_R28_00002** | | | |
| **Objekttitel:** | Fragment (Hand) einer Bronzeplastik aus Bregenz | | Motiv |
| **Objekttyp:** | Schwarz-Weiß-Fotografie | | Foto |
| **Technik/Material (Bildobjekt)** | • Fotografischer Abzug (Silbergelatineabzug | | Foto |
| **Klassifikation Motiv**: | Objekt | | Motiv |
| **Detailklassifikation Motiv** | Plastik | | |
| **Detailbeschreibung**: | Schwarz-Weiß-Fotografie eines Fragments einer Großbronzeplastik; linke Hand mit Unterarmansatz; Die Hand greift einen unbekannten Gegenstand; *hier darf noch etwas Text ,* XXXXX XXXXXXXXX; Material: Bronze, vergoldet; Fundort: Bregenz, Österreich; zeitliche Einordnung: römisch; Verwahrort d. Objekts: Vorarlberger Landesmuseum, Bregenz, Österreich; Foto: LEIZA; | | Verbindung aus Foto u Motiv |
| **Details Foto:** | 1 Foto) auf Bildkarte montiert; Karte mit Schreibmaschine beschriftet; | | |
| **Anzahl Foto:** | 1 | | Foto |
| **Bildnummer:** | LEIZA Neg.: T 80/2248 | | |
| | | | |

| Feld | Beispielbezeichnungen, Auswahlkategorien, | | Zuordnung |
|---|---|---|---|
| **Objekt O.41088** | | | |
| **Objekttitel:** | Frühmittelalterlicher Schnallenbeschlag aus Südspanien | | Motiv |
| **Objekttyp:** | Farb-Fotografie | | Foto |
| **Technik/Material (Bildobjekt)** | • Fotografischer Abzug<br>• Druck / Kopie<br><br>*(falls bekannt ergänze jeweils Detailbegriff. >> Liste)* | | Foto |
| **Klassifikation Motiv**: | Objekt | | Motiv |
| **Detailklassifikation Motiv** | Bekleidung | | |
| **Detailbeschreibung**: | Farbfotografie eines zungenförmigen Scharnierbeschlags mit paarigen Rundeln am Rand, verziert mit einem umlaufenden Kerbband und unkenntlichen gravierten Zeichen, darunter in der Mitte der Buchstabe A. Auf der Unterseite drei mittelständige Nietzapfen. Maße: L. 90 mm; B. 33 mm; <u>Material:</u> Bronze <u>Datierung</u>: 7. Jahrhundert n. Chr., <u>Fundort:</u> Südspanien; <u>Verwahrort d. Objekts:</u> LEIZA; LEIZA Inventarnummer: O.41088; | | Verbindung aus Foto u Motiv |
| **Details Foto:** | <u>Foto:</u> LEIZA; | | |
| **Bildnummer:** | LEIZA Neg.: T1990_00488-00489 | | |
| | | | |

| Feld | Beispielbezeichnungen, Auswahlkategorien, | | Zuordnung |
|---|---|---|---|
| **BK_R16_00013** | | | |
| **Objekttitel:** | Rekonstruktion eines römischen Schienenpanzers | | Motiv |
| **Objekttyp:** | Schwarz-Weiß-Fotografie | | Foto |
| **Technik/Material (Bildobjekt)** | • Fotografischer Abzug (Silbergelatineabzug) | | Foto |
| **Klassifikation Motiv**: | Objekt | | Motiv |
| **Detailklassifikation Motiv** | Militärische Geräte | | |
| **Detailbeschreibung**: | Schwarz-Weiß-Fotografie der Rekonstruktion eines römischen Schienenpanzers (Lorica segmentata); <u>Material:</u> Eisen, Bronze und Leder. *hier darf noch etwas Text* , <u>Fundort:</u> unbekannt; <u>zeitliche Einordnung:</u> römisch; <u>Datierung:</u> 90-100 n. Chr. Rekonstruktion und Geschenk von H. Russel Robinson (Tower of London, The Armouries); Rekonstruktion nach dem Hortfund aus dem Kastell von Corbridge am Hadrianswall, England, UK<u>; Verwahrort d. Objekts</u>: LEIZA, Mainz; <u>Verwahrort d. Original:</u> Museum of Antiquities, Newcastle-upon Tyne, UK. | | Verbindung aus Foto u Motiv |
| **Details Fotoobjekt:** | <u>Foto:</u> I. Feddersen/ LEIZA; 1 Foto) auf Bildkarte montiert; Karte mit Schreibmaschine beschriftet; | | |
| **Anzahl Foto:** | 1 | | Foto |
| **Bildnummer:** | LEIZA Neg. T 97/459; | | |
|  |  |  |  |

| Feld | Beispielbezeichnungen, Auswahlkategorien, | | Zuordnung |
|---|---|---|---|
| **BK_MM110_klasik_104300029** | | | |
| **Objekttitel:** | Athen, Theseion und Akropolis | | Motiv |
| **Objekttyp:** | Schwarz-Weiß-Fotografie | | Foto |
| **Technik/Material (Bildobjekt)** | • Fotografischer Abzug (Albumin) | | Foto |
| **Klassifikation Motiv**: | Architektur (architecture) | | Motiv |
| **Detailklassifikation Motiv** | Tempel | | |
| **Detailbeschreibung**: | Schwarz-Weiß-Fotografie einer Ansicht des Theseion und der Akropolis, Athen, Griechenland. Links im Bild ist das Theseion (auch *Hephaisteion* genannt) zu sehen, im Hintergrund die Akropolis; *hier darf noch etwas Text , … ;* <u>zeitliche Einordnung</u>: klassisches Griechenland; | | Verbindung aus Foto u Motiv (Datierung aus anderer Quelle) |
| **Details Foto:** | <u>Foto:</u> Fotograf und Herkunft des Fotos unbekannt; 1 Foto auf Bildkarte montiert; Karte handschriftlich beschriftet; Foto gelblich verfärbt; | | |
| **Anzahl Foto:** | 1 | | Foto |
| **Bildnummer** | - | | |

**Supplementary Information A1.2: Engineered text prompt used for *InternVL2***

To test the model's output quality without further fine-tuning of the model, the prompt had to be reformulated several times before the final descriptions could be generated by the model. The final prompt included detailed instructions on each category in the catalogue template and an example catalogue description was provided in the prompt. In addition, to avoid hallucinations, the prompt stressed to only use information given in the caption. Furthermore, the model was prompted in English since it was mostly pre-trained on English texts and hence performs best in English. The one English description selected from the three generated ones was translated into German and double-checked by a non-expert and an expert on typing errors and correct translations of archival terms.

Here is the engineered prompt which was entered with each labelled photo card:

Pretend to be an archivist who wants to catalogue this photo card digitally. Write a description including different fields which are provided below. Also use the text on the photo card.

The given image shows a photo card from the archive of the Leibniz Institute for Archaeology (LEIZA). It consists of a photo showing a motif and, if applicable, a German text. This text can often be found below the photo, sometimes as a structured template text listing fields like "FO." (Fundort, find spot), "Fdst." (Fundstelle, specific find spot), Kreis (county), Land (country), Museum (museum), "Neg." (Fotonegativ, photo negative), "Ggst." (Gegenstand, object name), "Zeit" (chronological classification) and "Lit." (relevant literature).

Task: Fill out the following fields with information directly extracted from the photo and the text. Only include information that is explicitly stated in the text.

Object title: What can be seen in the photo? If there is a structured template text on the photo card, use the main information from the field "Ggst" (object name) and "FO" (find spot).

Object type: Black and white photography / colour photography / graphic or drawing

Technique or material of the photo: Photographic print / print or copy / drawings (original) / paintings (original). Only add information in brackets if you are sure.

Here you can find the definition of the above techniques so you can distinguish them (especially photographic prints without a printing screen and prints with a printing screen):

Photographic prints are opaque photographs, usually positive (i.e., reproducing appearances without tonal reversal, otherwise use "negative prints"), usually on paper, and generally, but not always, printed from a negative). Photo prints do not have a printing screen, there are no visible dots. The photo paper is exposed to light, usually laser light, without a visible screen. The colours red, green and blue (RGB) are used.

Prints are pictorial works produced by transferring images by means of a matrix such as a plate, block, or screen, using any of various printing processes. Prints consist of small dots of ink, using the colours cyan, magenta, yellow and black (CMYK). By applying fine drops of ink in close proximity to the paper, all colours are mixed in the printed photo. On closer inspection, for example with a magnifying glass, these individual dots of colour are visible. This pattern is known as a print screen. So look closely at the photo and check if you can find a printing screen.

Drawings are visual works produced by drawing, which is the application of lines on a surface, often paper, by using a pencil, pen, chalk, or some other tracing instrument to focus on the delineation of form rather than the application of colour.

Paintings are unique works in which images are formed primarily by the direct application of pigments suspended in oil, water, egg yolk, molten wax, or other liquid, arranged in masses of colour, onto a generally two-dimensional surface.

Motif classification: Which of the following categories can be seen in the photo? Choose 1-2 categories: Settlements and landscapes / People / Architecture / Object

Here you can find some definitions:

Architecture: Structures or parts of structures that are the result of conscious construction, are of practical use, are relatively stable and permanent, and are of a size and scale appropriate for--but not limited to--habitable buildings. Works of architecture are manifestations of the built environment that is typically classified as fine art, meaning it is generally considered to have aesthetic value, was designed by an architect (whether or not his or her name is known), and constructed with skilled labor.

Objects: Broad classifications for material things that can be perceived with the senses, especially works of art and objects of historical material culture, works of art such as paintings, drawings, graphics, sculptures, decorative art, photographs and other cultural artefacts. An object can consist of one or more parts.

Detailed motif classification: Which motif is it exactly? Select 1-2 categories from the following categories (examples in brackets):

If Motif classification = Settlements and landscapes: Leave this field empty or add information you find in the text on the photo card.

If Motif classification = People: portrait / group picture

If Motif classification = Architecture: Monument / temple / arch of honour or triumphal arch / fortification / cityscape / circus / thermal baths / street / cult building / tomb / military complex / bridge / harbour / museum grounds / other

If Motif classification = Object: Sculpture (Statues, busts, reliefs, figures, statuettes, …) / art and painting (Painting, frescoes, mosaics, pictures, stucco, …) / vessels (Craters, amphorae, vases, plates, ...) / civilian equipment and tools (Spoon, knife, casting mould, ...) / military equipment (helmet, armour, lance, spear, sword, dagger) / furniture (e.g. fittings) / jewellery / coins and medals / clothing (fibula, brooch, belt buckles, …) / architectural elements (lighting, decor, inscription, brick, window, door, ...) / cult objects (amulet, …)

Detailed description: Describe what you see on the photo from left to right. Also use the text on the photo card, if applicable. Start with "Black and white photography / colour photography of a ..." Then write 1-2 keywords or short sentences. Write only information that you can find in the image or in the text!

Additional Information: If applicable, try to read the following sub-items from the text template on the photo card: Only if the photo shows an object: find spot ("FO." (Fundort), "Fdst." (Fundstelle, specific find spot), Kreis (county), Land (country)), chronological classification ("Zeit", e.g. Roman, Late Antique), dating (specific date), place of storage of the object ("Mus.", museum) and "Lit." (relevant literature).

Only if the photo shows settlements and landscapes: chronological classification ("Zeit"), location reference: Where was the photo taken?

Only if the photo shows people: Biographical information about the person: Does the text on the photo card give any information on the persons in the photo?

Only if the photo shows architecture: chronological classification ("Zeit"), location reference: Where was the photo taken?

Write only information that you can find in the image or in the text! If a sub-item is not described on the photo card, do not list the sub-item. Only use the information in the text on the photo card, if available. For locations, also include the city and state.

Photo details: Describe the following sub-items if they are described on the photo card: Photo (author and rights holder): If the author is not described on the photo card, do not list the sub-item.; Image carrier: X photo(s) mounted on photo card; Text: photo card is not labelled / labelled with typewriter / labelled by hand; photo is / is not discoloured (yellow). Look only at the photo when giving information on discolourings.

Number of photos: X

Image number:

If "Neg." (Fotonegativ, photo negative) appears in the text template of the photo card, enter the number after "Neg." here. Otherwise write "not known / specified". If there is a location indicated after "Neg.", write it in front of the number. Otherwise write "LEIZA" in front of the number.

Here is an example description written by an archivist:

Example 1:

Object title: Athens, Theseion and Acropolis

Object type: Black and white photography

Technique/material: Photographic print (albumen)

Motif classification: Architecture

Detailed motif classification: Temple

Detailed description: Black and white photograph of a view of the Theseion and the Acropolis, Athens, Greece. The Theseion (also known as the Hephaisteion) can be seen on the left, with the Acropolis in the background;

Additional information: chronological classification: Classical Greece;

Photo details: Photographer and origin of photo unknown; 1 photo mounted on photo card; card labelled by hand; photo discoloured yellow;

Number of photos: 1

Image number: not specified

Important Reminder: Fill out the fields with information directly extracted from the photo and the text. Only include information that is explicitly stated in the text!

*Supplementary Information A2. Model parameters*

In addition to the prompt engineering (Sahoo et al., 2024), several model parameters and sampling methods were tested on the *InternVL2* model, but only the *temperature* parameter showed clear and useful alternations in the output descriptions. The *temperature* parameter helps to determine how likely the model is to come up with different options for the next word. A temperature of zero means that the model is more deterministic and will always select the most probable next word. A temperature greater than zero involves a degree of randomness in the selection of the next word which makes the model's outputs less deterministic. In this work, a more deterministic model was favored since the information on the labelled photo card should not be altered but copied correctly into the corresponding data field of the catalogue description. However, a slightly higher temperature value could also reduce OCR errors since it lets the model generate words that are similar to the one that was falsely recognized on the labelled photo card. The *InternVL2* model thus generated three catalogue descriptions with different values of the *temperature* parameter (0.1, 0.5, 0.75) from which the best-fitting description was chosen by a human.

*Supplementary Information A3. Items presented in the experiment*

All texts in the experiment were presented in German. Their English translations can be found here.

For the independent variables:

- Archival expertise was measured (archive expert group / archie non-expert group)

    o as a categorical variable:

    "I am working as an archivist." (archive expert group) /

    "I am working as an archaeologist." (archive expert group) /

    "I am an apprentice in the archive school in Marburg." (archive expert group) /

    "I am studying archaeology, historical sciences or similar." (archive expert group) /

    "I research archives and collections as a hobby." (archive expert group) /

    "None of the above statements apply to me." (archive non-expert group)

    o as an ordinal variable – "How do you rate your expertise in archive practices and archival science?" (ranging from 1 = "very low" to 7 = "very high").

- Archaeological expertise was only measured on an ordinal scale – "How do you rate your expertise in the field of archaeology?" (ranging from 1 = "very low" to 7 = "very high").

- AI expertise was measured

    o as a categorical variable:

    "I have been working in the field of artificial intelligence, data science, computer science or similar for at least ten years (including my studies)." (AI expert group) /

    "My studies are in the field of artificial intelligence, data science, computer science or similar." (AI expert group) /

    "None of the above statements apply to me." (AI non-expert group)

    o as an ordinal variable – "How do you rate your expertise in the field of artificial intelligence?" (ranging from 1 = "very low" to 7 = "very high").

For the dependent variables:

a) Classification performance for AI-generated and expert-written descriptions – "Was this description generated by an AI? Yes / No", used for calculating (i) Signal Detection measures (sensitivity $d'$ and response bias $c$) and (ii) percentage correct responses.

b) Self-assessed classification performance – "You will now see 80 such descriptions. How many of them do you estimate you will correctly identify as AI-generated or not?

c) Perceived accuracy – "How many percent of this description is accurate?" (0-100%)

d) Perceived usefulness – "How many percent of this description is useful for scientific research?" (0-100%)

e) Willingness to use AI – "How frequently do you currently use artificial intelligence for you work and studies?" (ranging from 1 = "not very frequently" to 7 = "very frequently"; only measured in the pre-test, not used for data analysis) and "How likely are you to use artificial intelligence for your work or studies in a year?" (ranging from 1 = "not very likely" to 7 = "very likely", measured pre- and post-test)

f) Trust in AI – "I trust artificial intelligence in my work or studies." (ranging from 1 = "I completely disagree" to 7 = "I completely agree")

**Supplementary Information B. Results**

*Supplementary Tables B1*

**Supplementary Table B1.1: Predicted classification performance.**

|  | Predicted classification performance | |
|---|---|---|
| **AI-generated descriptions** | % correctly classified | 95% CI |
| Experts | .62 | [.56, .68] |
| Non-experts | .55 | [.52, .58] |
| **Expert-written descriptions** | | |
| Experts | .70 | [.63, .76] |
| Non-experts | .70 | [.66, .73] |

*Note.* Degrees of freedom were *Df* = 137.

**Supplementary Table B1.2: Correlations of self-assessed expertise with self-assessed classification performance (H2c)**

|  | Correlation with self-assessed classification performance | |
|---|---|---|
| **Self-assessed expertise in** | *r* | *p* |
| Archives | -.04 | .662 |
| Archaeology | .08 | .351 |
| Artificial Intelligence | .11 | .208 |

*Note.* Degrees of freedom were *Df* = 137.

**Supplementary Table B1.3: Correlation of accuracy and usefulness ratings and their comparison**

|  | r | p |
|---|---|---|
| **AI-generated descriptions** |  |  |
| Experts | .87 | < .001 |
| Non-experts | .69 | < .001 |
| **Expert-written descriptions** |  |  |
| Experts | .90 | < .001 |
| Non-experts | .72 | < .001 |

*Note.* Degrees of freedom were $Df_{experts}$ = 27 and $Df_{nonexperts}$ = 108. Correlations were significantly different both for ratings of AI-generated descriptions, $z$ = 2.20, $p$ = 0.14, and of expert-written descriptions, $z$ = 2.60, $p$ = .009.

**Supplementary Table B1.4: Accuracy and usefulness ratings**

|  | *M* | *SD* |
|---|---|---|
| **Accuracy ratings** |  |  |
| AI-generated | 67.61 | 12.92 |
| Expert-written | 76.11 | 12.31 |
| **Usefulness ratings** |  |  |
| AI-generated | 58.01 | 14.75 |
| Expert-written | 70.01 | 14.33 |

*Note.* Means and standard deviations of accuracy and usefulness ratings in %. Question items can be found in Supplement A3.

**Supplementary Table B1.5: Expert-specific willingness to use AI and trust in AI**

|  | pre | | post | |
|---|---|---|---|---|
|  | *M* | *SD* | *M* | *SD* |
| **Willingness to use AI tools** | | | | |
| Experts | 4.66 | 1.82 | 4.41 | 1.68 |
| Non-experts | 5.64 | 1.52 | 5.27 | 1.61 |
| **Trust in AI tools** | | | | |
| Experts | 3.48 | 1.02 | 3.41 | 1.21 |
| Non-experts | 3.96 | 1.30 | 3.73 | 1.44 |

*Note.* Means and standard deviations of willingness to use AI and trust in AI, reported on a seven-point Likert scale. Question items can be found in Supplement A3.

---

[i]

https://zenodo.org/records/15640284?preview=1&token=eyJhbGciOiJIUzUxMiJ9.eyJpZCI6ImFmMzBlNDE3LWY1NGQtNDBjZi1iMzM5LTM1Y2QxMWQxOGZmMCIsImRhdGEiOnt9LCJyYW5kb20iOiI2MDlhZTBlMGNhODE4ZDUwYjIyYTVkODY4MzU5N2I0YSJ9.KOoC9URvTnu2Zk1Q4TTSKf5_9twIwSSlP46xB4xksrXZeTMnlf3Zf5gNsRhe2AvajRNkdlKX1hgJk6rsNDyuUw

[ii]

https://zenodo.org/records/15640284?preview=1&token=eyJhbGciOiJIUzUxMiJ9.eyJpZCI6ImFmMzBlNDE3LWY1N

GQtNDBjZi1iMzM5LTM1Y2QxMWQxOGZmMCIsImRhdGEiOnt9LCJyYW5kb20iOiI2MDlhZTBlMGNhODE4ZDUwYjIyYTVkODY4MzU5N2I0YSJ9.KOoC9URvTnu2Zk1Q4TTSKf5_9twIwSSlP46xB4xksrXZeTMnlf3Zf5gNsRhe2AvajRNkdlKX1hgJk6rsNDyuUw